# Impact of Nanometer-Thin Stiff Layer on Adhesion to Rough Surfaces

Shubhendu Kumar[1], Babu Gaire[1], Bo Persson[2,3], and Ali Dhinojwala[1*]

[1]School of Polymer Science and Polymer Engineering, The University of Akron, Akron, OH 44325 USA.

[2]Peter Grünberg Institute (PGI-1), Forschungszentrum Jülich, Jülich 52425, Germany

[3]Multiscale Consulting, Wolfshovener Str. 2, Jülich 52428, Germany

[*]Corresponding author. Email: ali4@uakron.edu

## Abstract

Adhesion requires molecular contact, and natural adhesives employ mechanical gradients to achieve complete (conformal) contact to maximize adhesion. Intuitively, one expects that the higher the modulus of the top layer, the lower will be the adhesion strength. However, the relationship between the thickness of the stiff top layer and adhesion is not known. In this work, we quantified the adhesion between a stiff glassy poly (methyl methacrylate) (PMMA) layer of varying thickness on top of a soft poly dimethyl siloxane (PDMS) elastomer with a sapphire lens. We found that only $\approx$ 90 nm thick PMMA layer on a relatively thicker, softer, and elastic PDMS block is required to drop macroscopic adhesion to almost zero during the loading cycle. This drop in adhesion for bilayers can be explained using a conformal model developed by Persson and Tosatti, where the elastic energy to create conformal contact depends on both the thickness and the mechanical properties of the bilayer. A better understanding of the influence of mechanical gradients on adhesion will have an impact on adhesives, friction, and colloidal and granular physics.

## Introduction

Adhesion between two surfaces requires molecular contact, and even a small roughness disrupts this molecular contact; this phenomenon is commonly known as the *adhesion paradox* [1–3]. Surfaces with low modulus can deform under pressure to create molecular contact [4,5]. However, achieving conformability requires additional energy for elastic deformation, which reduces the work of adhesion during approach [6]. When this additional elastic energy exceeds the intermolecular work of adhesion, a complete loss of macroscopic adhesion is observed, and the actual (or nominal) contact area is then dominated by Hertzian mechanics [7]. The mathematical formulation for calculating the effective work of adhesion was proposed by Persson and Tosatti in 2001 using height power spectral density (PSD), and this model has been validated for soft elastomeric siloxane polymers in contact with hard diamond surfaces by Dalvi et al. [8,9].

Many natural and man-made surfaces have gradient mechanical properties [10–12]. These mechanical gradients may result from differences in the composition or topological differences that result in an effective modulus that is a function of thickness [10–14]. For example, geckos have setae that branch into finer spatula and produce mechanical gradients controlled by both chemical composition and differences in physical parameters such as the diameter and length of the spatula as compared to setae [13,15]. These chemical or mechanical gradients can range from nanometer to micrometer scale and, in some cases, they help or deter mechanical contact. Such gradients offer an exquisite approach for controlling molecular contact and, thus, adhesion and friction. [16]

Here we address the fundamental question of how a mechanical gradient can affect adhesion in contact with rough surfaces. To simplify the system, we considered a bilayer system with a high-modulus, glassy polymer poly (methyl methacrylate) (PMMA) on top of a soft elastomer poly dimethyl siloxane (PDMS), where the thickness of PMMA ranged between 10 and 90 nms. We measured the adhesion using the Johnson–Kendall–Roberts (JKR) technique by bringing the bilayer into contact with a sapphire hemispherical lens during approach and retraction. We compared the experimental results with predictions of the Persson-Tosatti model modified for bilayers [17]. An understanding of how adhesion is affected by mechanical gradients will shed light on the design principles of biological structures and inform the design of effective adhesives that can improve adhesion by controlling the effective modulus of the topmost layer in many applications including robotics and the biomedical sciences [12,16,18].

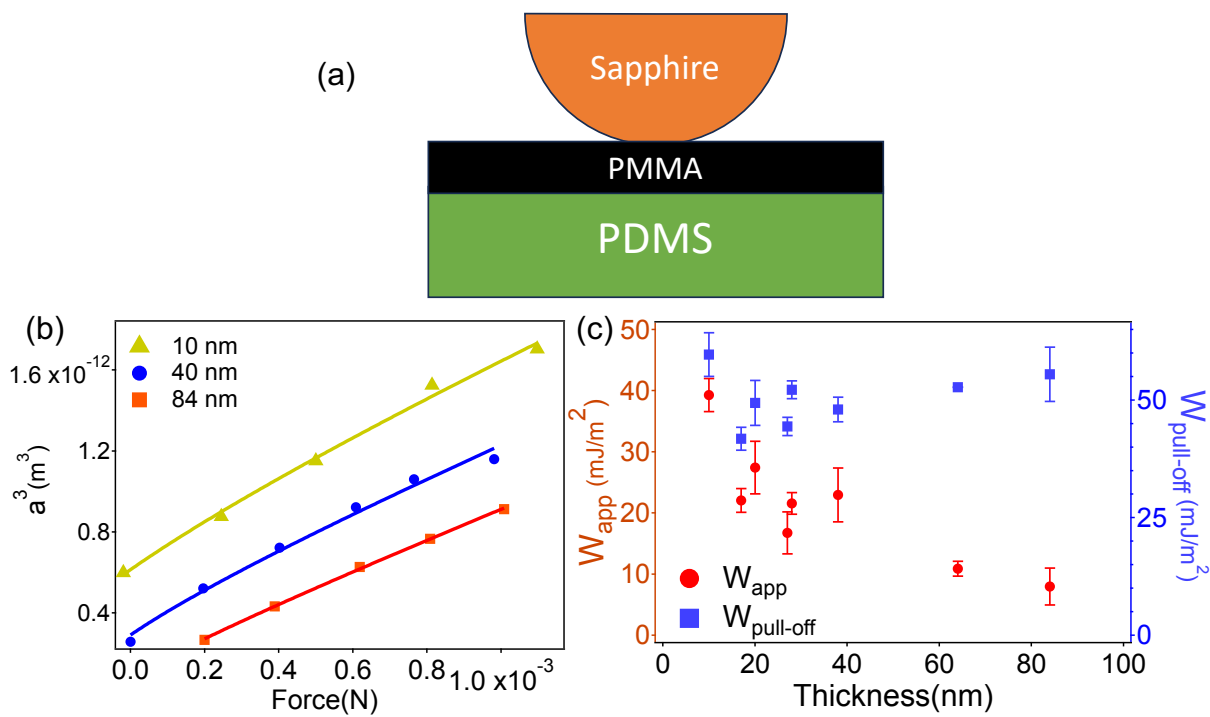

**Figure 1:** *(a) The geometry showing a sapphire lens bought in contact with a bilayer sample consisting of a top layer of rigid/glassy PMMA on a soft PDMS elastomer. The lens in contact with a flat sheet is used for JKR experiments in which the contact area is measured as a function of applied load. (b) Contact radius cubed vs. applied force data for bilayer samples having PMMA layers with different thicknesses. The solid line is fit*

*using the JKR contact mechanics model. (c) Plot of adhesion values obtained from JKR fits during approach ($W_{app}$; shown as red symbols). Pull-off ($W_{pull-off}$) values were determined by retracting the sapphire lens with a speed of 60 nm/s and using the JKR analysis (shown by blue symbols).*

## Results and Discussion

**Figure 1** summarizes the adhesion results for bilayers with PMMA thickness in the range of 10–84 nanometers. **Figure 1A** illustrates the JKR geometry used to measure the bilayer adhesion. A sapphire lens with a radius of curvature of 1.25 mm is brought into contact with the bilayer at a speed of 60 nm/s and the contact area as a function of load is measured in a semi-static way; the results are shown in **Figure 1B** for 10-, 40- and 84-nm-thick PMMA layers. The solid lines in **Figure 1B** are the fit using the JKR equation to determine the apparent work of adhesion ($W_{app}$), and the effective moduli determined from the fits are summarized in Table S1 (in SI). During retraction, we measured the pull-off force and, using the JKR equation, we related the pull-off forces to the work of adhesion during pull-off ($W_{pull-off}$). Both of these values are plotted as a function of film thickness in **Figure 1C**. The values of $W_{app}$ drop rapidly as we increase the thickness (Pearson correlation coefficient of –0.84). In comparison, the values of $W_{pull-off}$ are not affected by PMMA thickness (Pearson correlation coefficient of 0.27).

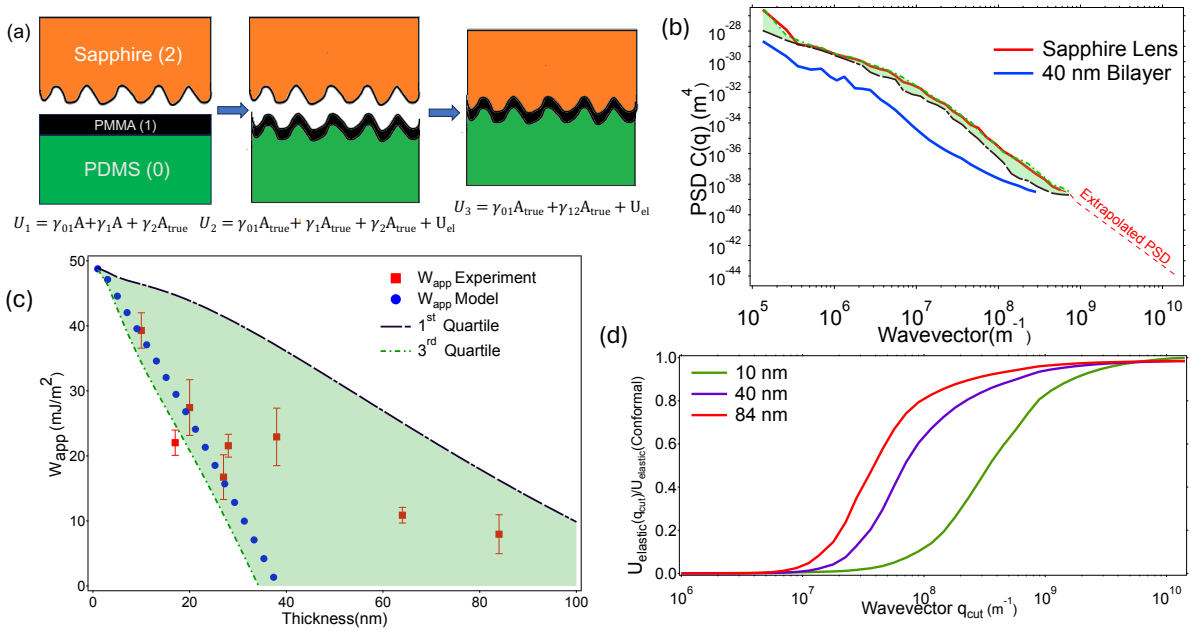

*Figure 2: (a) Schematic showing different stages of a bilayer sample as it conforms to the roughness of the solid sapphire substrate. The center panel shows the intermediate stage to illustrate that the bilayer must deform to match the contours of the rough surface. The final panel shows a conformal contact with the rough surface. The equations provide the total energy of the system in each stage and the apparent work of adhesion is the difference between the final and the initial stage. (b) Results for the PSD (C) of the*

sapphire lens and a 40-nm bilayer system. The shaded region represents variation in PSD of the sapphire lens within the interquartile range where the green dotted line represents 3$^{rd}$ quartile and the black dotted line represents the 1$^{st}$ quartile. The dotted red line represents the extrapolated average (C) of the sapphire lens. These profiles were measured using an atomic force microscope, and PSD was calculated using a previously published procedure [19,20]. (c) The experimental adhesion values for the bilayers in contact with sapphire are plotted as a function of thickness and are compared with the theoretical values predicted using **Equations 1–5**. The blue circles used the average values of C, and the shaded regions are calculated using the interquartile range of C (first and third quartile). (d) The ratio of elastic energy spent by several bilayer systems as a function of $q_{cut}$. These values are normalized by $U_{elastic}$ after integrating **Equation 2** with $q_1 = 10^{10}$ (m$^{-1}$).

To determine $W_{app}$ for the bilayers, the original Persson–Tosatti formalism was modified to account for changes in surface/interfacial energies of the PMMA ($\gamma_1$) and PMMA–PDMS ($\gamma_{01}$) interfaces. In addition, we needed to account for the effective elastic energy to conform to the roughness of the solid surface [8]. **Figure 2A** shows the steps involved in the development of the model to calculate $W_{app}$. Here, $A_{app}$ is the projected area, $\gamma_2$ is the surface energy of the rough solid surface, and $U_{el}$ is the elastic energy to deform the bilayer to conform to the rough solid surface. $W_{app}$ can be calculated based on the energy difference to separate the two surfaces.

$$W_{app} = W_{int} A_{true}/A_{app} - (\gamma_1 + \gamma_{01})\left(A_{true}/A_{app} - 1\right) - U_{el}/A_{app} \quad (1)$$

To determine the change in surface energy in **Equation 1** due to the 0-1 interface, we have assumed that the 0-1 interface has the same PSD as the 1-2 interface after conformal contact, independent of film thickness. This is strictly true only for long wavelength roughness with the wavelength $\lambda \gg d$ (where $d$ is the film thickness). For the bilayers studied here, $\gamma_{01}$ is small and the accuracy of this assumption does not influence the adhesion values. The elastic energy can be determined using the PSD of the rough surface, C (the definition of C is based on reference [21]).

$$U_{el}/A_{app} = E_1^* \pi/2 \int_{q_0}^{q_1} dq\, q^2 \frac{C(q)}{S(q)} \quad (2)$$

In **Equation 2**, $E_1^* = E_1/(1-v_1^2)$ is the modulus of layer **1 (top layer)** and $S(q)$ is the dimensionless surface responsive function, which depends on $qd$. The $q$ in **Equation 2** is the wavevector (which is equal to $2\pi/\lambda$, where $\lambda$ is the wavelength of roughness), and $d$ is the thickness of the top layer. $S(q)$ also depends on the Poisson ratio of two layers ($v_o$ and $v_1$) and the ratio of Young's modulus of the top and bottom layer ($E_1/E_0$) [17]

$$S(q) = \frac{1 + 4mqde^{-2qd} - mne^{-4qd}}{1 - (m+n+4mq^2d^2)e^{-2qd} + mne^{-4qd}} \quad (3)$$

The parameters $m\ and\ n$ are expressed as follows:

$$m = \frac{\mu_1/\mu_0 - 1}{\mu_1/\mu_0 + 3 - 4v_1} \tag{4}$$

and

$$n = 1 - \frac{4(1-v_1)}{1 + (\mu_1/\mu_0)(3 - 4v_0)} \tag{5}$$

where the shear moduli are related to Young's modulus: $\mu_0 = E_0/(2(1 + v_0))$ and $\mu_1 = E_1/(2(1 + v_1))$.

The value of $S(q)$ as a function of $q$ for PMMA/PDMS in contact with the sapphire substrate is provided in **Figure S1** (SI). As $d$ tends to zero, **Equation 1** will converge to the value expected for a PDMS (layer 0) in contact with a rough substrate. As $d$ tends to infinity, **Equation 1** will converge to the expected value for PMMA (layer 1).

To predict $W_{app}$ for the system, we need to measure $C$ for the sapphire lens and the PMMA–PDMS bilayer. **Figure 2B** shows the PSD for these two surfaces and the effective PSD is a function of the sapphire lens, since the bilayers are much smoother in comparison. The other parameters used to calculate $W_{app}$ using **Equations 1-5** are provided in **Table S2** (SI). The only unknown $W_{int}$ is the intrinsic work of adhesion between the PMMA layer and the sapphire lens. The value for $W_{int}$ of 49 mJ/m$^2$ was determined by minimizing the least squared error. Recent studies suggest that the modulus of PMMA could be a function of thickness. Accounting for this results in slightly higher values of adhesion for lower thickness and those results are summarized in **Figure S3** (SI). [22,23]

The theoretical predictions of $W_{app}$ are shown in **Figure 2C** as a function of thickness (circles). The contribution of the surface area terms and the elastic energy in **Equation 1** are plotted in **Figure S2** (SI). The second term is much smaller in magnitude because the area ratio term ($A_{true}/A_{app}$) is close to 1 for these almost smooth sapphire lenses. When the elastic contribution exceeds $W_{int}$, the adhesion drops to zero. The initial drop in $W_{app}$ measured experimentally matches well with the theoretical predictions for smaller thicknesses. However, the $W_{app}$ from the experiment dropped off more slowly than predicted by theory. One reason for these discrepancies may be due to the extreme sensitivity of $W_{app}$ to roughness. For example, if we take the standard deviation based on different measurements of PSD of the sapphire lens and calculate $W_{app}$, we observe the predictions are spread out in the shaded region predicted by the bilayer model. This finding is intriguing because the results suggest that the sensitivity to roughness is a function of the thickness of the PMMA layer. Another reason for a slower drop in adhesion could be due to plastic deformation of PMMA film.

In **Figure 2D** we plot the ratio of $U_{elastic}(q_{cut})/U_{elastic\ (total)}$ as a function of $q_{cut}$. For this calculation, we had to extrapolate the PSD to higher $q$ values beyond the limit we measured experimentally. We found that for thinner film, this ratio converges to 1.0 at higher $q$ values than that for a thicker film. Therefore, the variation in the roughness of the sapphire lens at lower $q$ values (or larger wavelengths) affects the thicker film more than the thinner film, producing higher standard deviations. However, identifying specific reasons for deviation for thicker PMMA films will require a much more uniform rough surface with very little variation in roughness at lower $q$ values.

The application of **Equation 1** to explain the adhesion results assumes that for all these samples, the contacts are conformal. We used surface-sensitive infrared-visible sum frequency generation (SFG) spectroscopy to address this question. Using SFG we measured the shift of the vibrational peak of the OH groups on the sapphire surface after contact with the PMMA surface using a bilayer geometry. This shift is due to hydrogen bonding between surface OH groups and carbonyl groups in PMMA, which is sensitive to variation in separation of less than 0.3-0.4 nm [24,25]. We observed no differences in the shift of the SFG peak for surface OH between the 10- or 200-nm-thick PMMA bilayers and no differences as a function of the applied load. The SFG results are summarized elsewhere [26]. This confirms that in these thickness ranges, the contact is conformal—and that **Equation 1** is applicable for modeling the adhesion data.

Finally, we explain the insensitivity of $W_{pull\text{-}off}$ to the thickness of the PMMA layer. We observed that for uniform PDMS layers in contact with rough surfaces, $W_{pull\text{-}off}$ does not follow the conformal adhesion model (**Equation 1**) but is determined by contact line pinning and can be explained using the Griffith model, where the adhesion energy is the product of the real contact area and the $W_{int}$. The $W_{int}$ value for all these thicknesses is dictated by PMMA–sapphire interaction, and they are similar for all these measurements. The long-range van der Waals interaction is almost independent of the thickness of the PMMA layer. (**Figure S4**, SI) [27,28]. Based on our SFG results, we concluded that these samples are conformal and, since both $W_{int}$ and the real contact area are not a function of the thickness of the PMMA layer, we expect $W_{pull\text{-}off}$ to be a constant, in agreement with our experimental results.

In summary, we show that adhesion is extremely sensitive to the bilayer thickness in contact with solids of modestly low roughness. In the short range in thickness between 10 and 90 nm, the adhesion during approach is lost due to roughness, as the elastic energy to conform to roughness is comparable to the intrinsic thermodynamic work of macroscopic adhesion. The predictions using a conformal bilayer contact model compare well with the experimental results and point towards extreme sensitivity of adhesion to small differences in roughness. These results provide insights on the control of adhesion by controlling the near-surface chemistry and modulus. For example, the model presented here can also predict how adhesion can be improved in contact with rough surfaces by layering a thin film of low-modulus polymers on a rigid polymer underlayer. This finding has a direct impact in many fields including soft robotics as well as adhesives for engineering and biomedical applications.

**Acknowledgments**

The authors acknowledge financial support for this work from the National Science Foundation (DMR-2208464). AD also acknowledges the financial support from the Knight Foundation (W. Gerald Austen Endowed Chair). The authors wish to thank Prachi Karanjkar, Dr. Felix Cassin (University of Pittsburgh), and Professor Tevis Jacobs (University of Pittsburgh) for their help with AFM measurements. In addition, we thank Utkarsh Patil, Dr. Stephen Merriman, and Dr. Nityanshu Kumar for their helpful comments and discussion.

# Supplementary Materials

# Impact of Nanometer-Thin Stiff Layer on Adhesion to Rough Surfaces

Shubhendu Kumar[1], Babu Gaire[1], Bo Persson[2,3], and Ali Dhinojwala[1*]

[1]School of Polymer Science and Polymer Engineering, University of Akron, Akron, OH 44325 USA.

[2]Peter Grünberg Institute (PGI-1), Forschungszentrum Jülich, Jülich 52425, Germany

[3]Multiscale Consulting, Wolfshovener Str. 2, Jülich 52428, Germany

[*]Corresponding author. Email: ali4@uakron.edu

## Supplementary Information

### Materials and Substrates:

Poly(methyl methacrylate) (PMMA) with a molecular weight of 100,000 g/mol and a polydispersity of 1.09 was purchased from Scientific Polymer Products (Ontario, NY, USA). Different concentrations of PMMA samples were dissolved in toluene and spun coated to prepare PMMA films of thickness ranging from 10 to 84 nms. These films were dried at room temperature for 48 hours before preparing the bilayer samples. Sheets of polydimethylsiloxane (PDMS) were prepared based on the procedure described by Dalvi et al. [1]. PDMS sheets were Soxhlet-extracted using toluene at 124 °C for 72 h to remove any unreacted PDMS. After 24 h of drying in air, the PDMS sheets were dried under vacuum at 120 °C overnight to remove any remaining toluene. The PMMA films were transferred on top of the PDMS sheets using a previously reported film transfer procedure [2]. The PMMA films adhere to PDMS by van der Waals interactions. Hemispherical sapphire lenses (2.5 mm in diameter) were purchased from Edmund Optics. All chemicals were used as received.

### Adhesion Measurement:

The adhesion experiment was done using the custom-built Johnson–Kendall–Roberts (JKR) test setup, in which a sapphire lens of diameter 2.5 mm was brought in contact with the bilayer system as shown in **Figure 1(a)**. The load and contact area were measured simultaneously until the maximum load of 1 mN. The apparent work of adhesion ($W_{app}$) was calculated by fitting contact area versus force data to the JKR equation (fitting results shown in **Figure 1 (b)**). Once the maximum load was applied, the sapphire lens was retracted at the same speed, and the pull-off force was recorded. The pull-off force was then used to calculate the pull-off work of adhesion ($W_{pull-off}$). [3]

### Roughness Measurement:

Roughness for the spherical lens and the 40-nm bilayer film were measured by atomic force microscopy (AFM) using a Bruker Dimension Icon microscope. For the spherical lens, a total of 28 scans with scan sizes of 100 nm, 250 nm, 500 nm, 750 nm, 1 µm, 5 µm 10 µm, 25 µm and 50 µm were collected using Tap DLC 150 tips, which have a radius of less than 15 nm, a force constant of 5 N/m, and a resonance frequency of 150 kHz. A different tip, a RTESPA 300 tip with a nominal tip radius of 8 nm was used for measuring the roughness of the bilayer films. The AFM images were uploaded to a website known as Contact Engineering (https://contact.engineering) to calculate the one-dimensional power spectra density (PSD), which was later converted to a two-dimensional PSD and then to a Persson's PSD ($C$) using the equations provided in **Reference 4** [1,4,5]. The $C$ data were used to calculate the elastic energy using **Equation 2** (main text) [4].

**Table S1:** Experiment work of adhesion and modulus data obtained by fitting the JKR equation:

$$(a^3 = {3R}/{4E^*}\left(F + 3\pi R W_{app} + \sqrt{6\pi R F W_{app} + (3\pi R W_{app})^2}\right))$$

for the thicknesses of PMMA film on a thick PDMS layer. In the JKR equation, $R$ is the radius of curvature, $F$ is force and a is the contact radius.

| Thickness (nm) | $W_{app}$ (mJ/m²) | $E^*$ (Effective Young's Modulus (MPa) of the bilayer in contact with sapphire) |
|---|---|---|
| 10 | 39.3±2.7 | 1.9±0.02 |
| 17 | 22.0±1.9 | 2.0±0.05 |
| 20 | 27.4±4.3 | 1.8±0.05 |
| 27 | 16.7±3.4 | 1.8±0.13 |
| 28 | 21.6±1.8 | 1.8±0.13 |
| 39 | 22.9±4.4 | 2.0±0.04 |
| 64 | 10.9±1.2 | 2.2±0.15 |
| 84 | 8.0±3.0 | 2.1±0.18 |

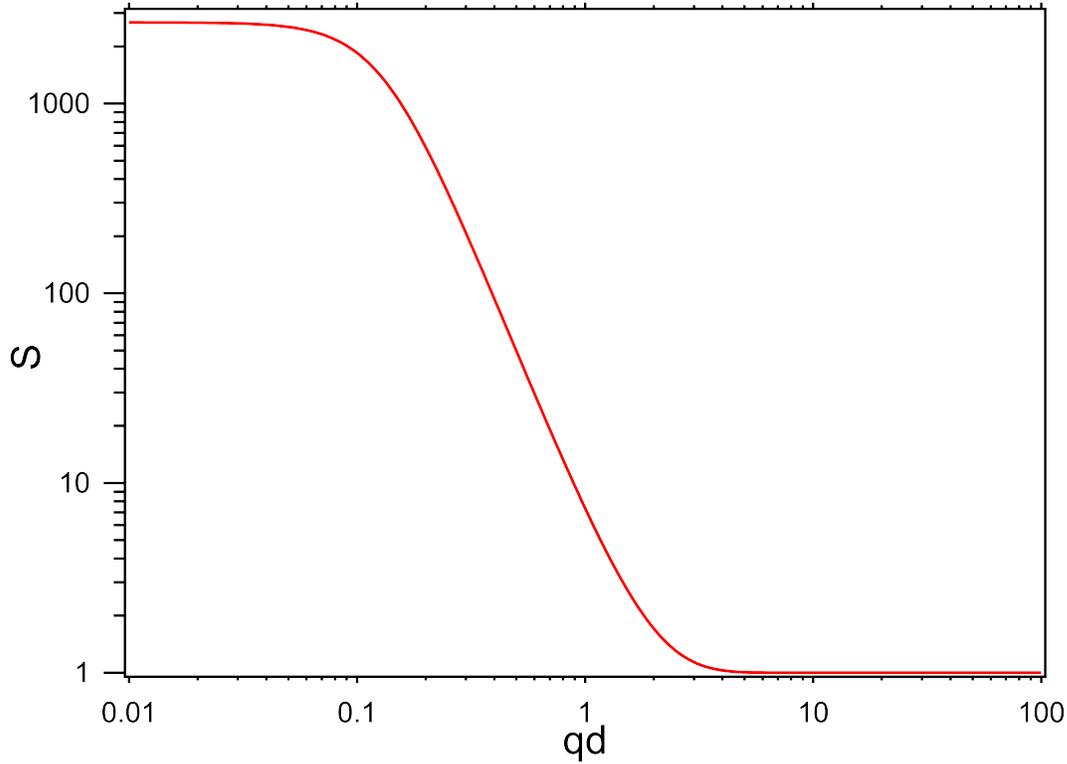

**Figure S1**: Dimensionless parameter $S(qd)$ vs. $qd$ using the geometry of PMMA thin layer on top of a thicker PDMS film. [6]. As the thickness of the PMMA film tends to infinity, the $S(qd)$ converges to 1. As $d$ converges to zero, the value of $S(qd)$ converges to the ratio $\frac{E_1}{1-v_1^2} / \frac{E_0}{1-v_0^2}$

**Table S2**: Parameters used to calculate theoretical work of adhesion during the approach cycle.

| Parameters | Value | Reference |
| --- | --- | --- |
| $W_{int}$ Intrinsic work of adhesion | 49 mJ/m² | -- |
| $\gamma_1$ Surface energy of PMMA | 39 mJ/m² | [7] |
| $\gamma_{01}$ Interfacial energy of PDMS-PMMA | 4 mJ/m² | [7,8] |
| $E_1$ Modulus of PMMA | 3 GPa | [7,9] |
| $E_o$ Modulus of PDMS | 1 MPa | [1] |
| $v_1$ Poisson ratio of PMMA | 0.4 | [7,10] |
| $v_o$ Poisson ratio of PDMS | 0.5 | [1] |

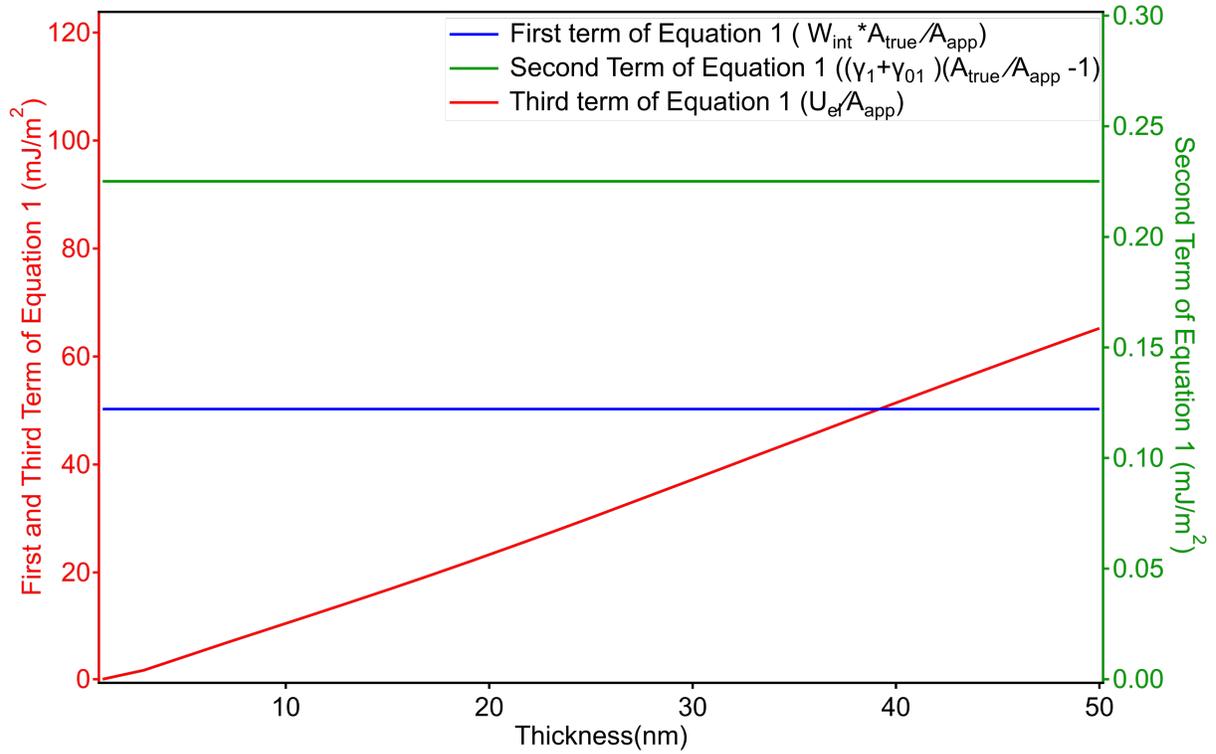

**Figure S2:** Comparison of the magnitude of different components contributing to $W_{app}$ in **Equation 1** (main text). The first ($W_{int}$) and the third term ($U_{elastic}$) are dominant and are shown using the left y-axis. Since the area ($A_{true}/A_{app}$) ratio is close to 1, the second term is much smaller and is shown using the right y-axis.

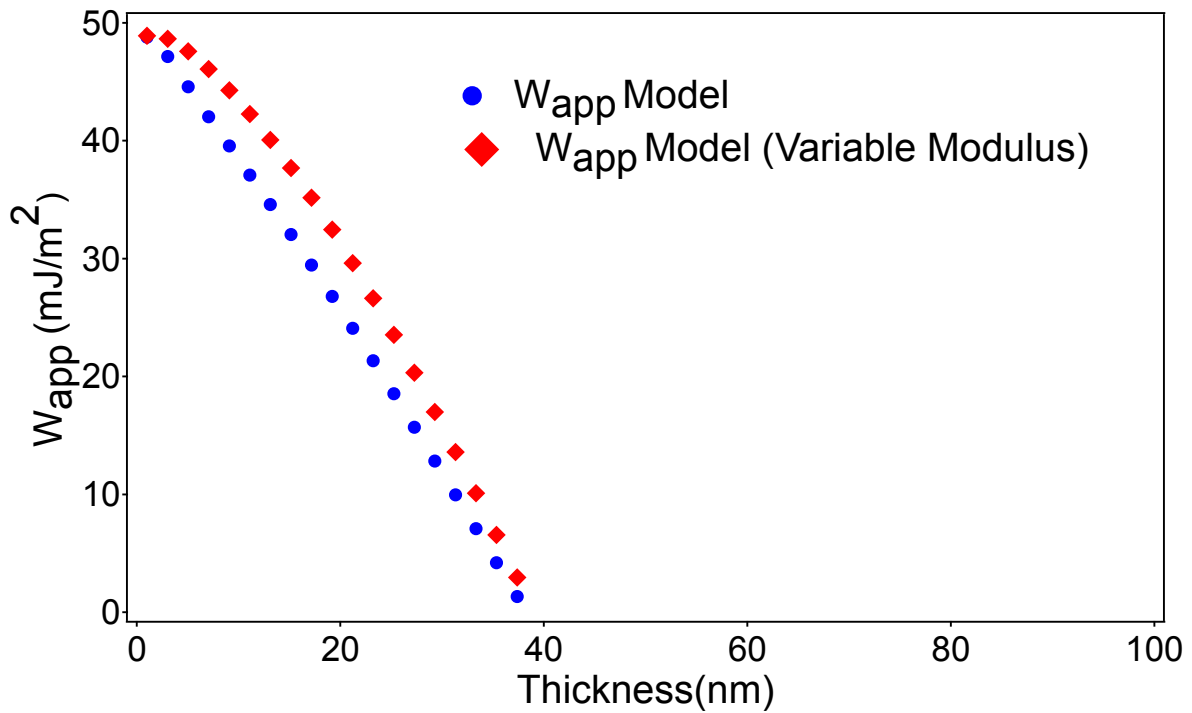

**Figure S3:** Comparison of the calculated $W_{app}$ value using **Equation 1** (main text) when assuming a uniform modulus across all thicknesses (represented by the blue symbols) against the scenario where the modulus changes with film thickness (illustrated by the red symbols).

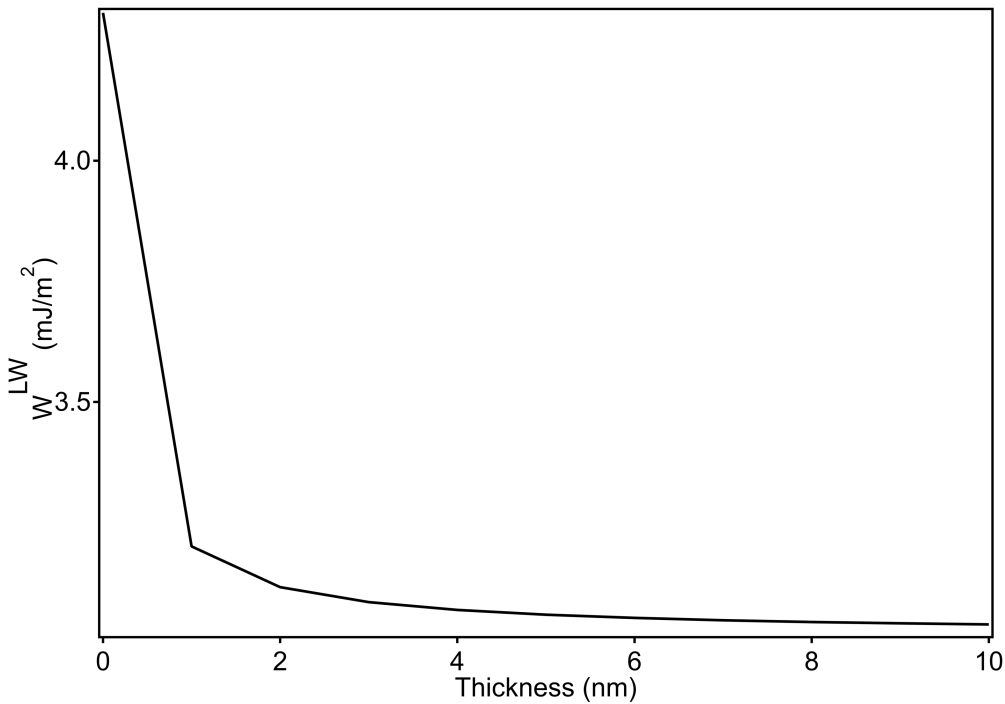

**Figure S4**: Variation in the Lifshitz–van der Waals contributions to the adhesion energy as a function of the thickness of the bilayer with PMMA layer on top of a much thicker PDMS film. [11,12] For the thickness of the samples used in the experiments, the contribution of $W^{LW}$ is independent of thickness.